\documentstyle[psfig,aasms4]{article}
\slugcomment{accepted by Astronomical Journal, April 1, 2003}
 
\begin{document}
 
\title{Sensitive observations at 1.4 and 250 GHz of $z~ > ~5$ QSOs}
\author{A. O. Petric}
\affil{Astronomy Department, Columbia University, New York, NY USA \\
andreea@astro.columbia.edu}
\author{C. L. Carilli}
\affil{National Radio Astronomy Observatory, P.O. Box O, Socorro, NM,
87801, USA \\
ccarilli@nrao.edu}
\author{F. Bertoldi}
\affil{Max-Planck Institut f\"ur Radioastronomie, Auf dem H\"ugel
69, Bonn, D-53121, Germany}
\author{Xiaohui Fan}
\affil{Steward Observatory, Univ. of Arizona, Tucson, AZ, 85721}
\author{P. Cox}
\affil{Institut d'Astrophysique Spatiale, Universit\'e de Paris XI,
91405 Orsay, France}  
\author{Michael A. Strauss}
\affil{Princeton University Observatory, Peyton Hall, Princeton,
NJ 08544, USA}
\author{A. Omont}
\affil{Institut d'Astrophysique de Paris, CNRS, 98 bis boulevard
Arago, F-75014, Paris, France} 
\author{Donald P. Schneider}
\affil{Dept. of Astronomy, Pennsylvania State University, 
University Park, PA 16802, USA} 

\begin{abstract}

We present 1.4 and 5~GHz observations taken with the Very Large Array
(VLA), and observations at 250~GHz obtained with the Max-Planck
millimeter bolometer (MAMBO) at the IRAM 30~m telescope, of ten
optically selected Quasi-stellar Objects (QSOs) at {5.0~$\leq
~z~\leq~6.28$}. Four sources are detected at 1.4 GHz.  Two of the
sources have rest-frame 1.4 GHz luminosity densities $> 5.0~ \times
~10^{26}$~W Hz$^{-1}$, placing them in the regime of radio-loud
QSOs. Both of these sources are also detected at 5~GHz. These results
are roughly consistent with there being no evolution of the radio-loud
QSO fraction out to $z \sim 6$.

Three sources have been detected at 250 GHz or 350 GHz by these, and
previous, observations. The (sub)mm flux densities for these three
sources are much larger than their 1.4 GHz flux densities. The rapidly
rising spectra into the (rest frame) far IR argue that the observed
mm emission is likely thermal emission from warm dust, although more
exotic possibilities cannot be precluded.  The implied IR luminosities
are between $10^{12}$ and 10$^{13}$ L$_\odot$.  For J0301+0020 the
radio continuum emission is clearly above that expected for a star
forming galaxy based on the radio-FIR correlation.  In this case it
seems likely that the radio emission relates to the AGN.  For
J0756+4104 the radio emission is within the range expected for a star
forming galaxy, while for J1044--0125 the radio upper limit is at
least consistent with a star forming galaxy.  If the dust is heated by
star formation, the implied massive star formation rates are between
200 and 1000 M$_\odot$ year$^{-1}$.

We do not detect radio emission from the reported X-ray jet associated
with J1306+0356. The lack of radio emission implies that the magnetic
field is well below typical equipartition values in powerful radio
jets, or that particle acceleration ceased between $10^6$ and $10^7$
years ago, or that the X-ray emission is not inverse Compton emission
from a jet related to J1306+0356.

The highest redshift source in our sample (J1030+0524 at $z=6.28$) is
not detected at 1.4 or 250 GHz, but four fairly bright radio sources
($\rm S_{1.4GHz} > 0.2$mJy) are detected in a 2\arcmin~ field
centered on the QSO, including an edge-brightened ('FRII') double
radio source with an extent of about 1\arcmin. 
A similar over-density of radio sources is seen in the field
of the highest redshift QSO J1148+5251.  We speculate that
these over-densities of radio sources may indicate  clusters along the
lines-of-sight, in which case gravitational lensing by the cluster could
magnify the  QSO emission by a factor 2 or so without giving rise
to  arcsecond-scale distortions in the optical images of the QSOs.

\end{abstract}
 

\section{Introduction}

Surveys such as the Sloan Digital Sky Survey (SDSS) (York et al. 2000)
have generated large samples of QSOs out to {z$\sim$6.}  Optical
spectroscopy of two of the highest z QSOs revealed, for the first
time, evidence for broad, optically thick absorption regions in the
(rest frame) far UV, as expected for the neutral intergalactic medium
(Gunn \& Peterson 1965; Becker et al. 2001; Djorgovski et al. 2001;
Pentericci et al. 2002; Fan et al. 2003), suggesting that we are
probing into the epoch of reionization, the edge of the ``dark ages''
when the first stars and massive black holes were formed.  Assuming
that the intrinsic continuum spectra of the {z~$>$~5} QSOs are similar
to those of nearby QSOs as compiled by Elvis et al. (1994), and that
these QSOs are not lensed, Eddington limit arguments result in lower
limits of several times $10^9~\rm{M_{\odot}}$ for the masses of the
black holes in these high redshift objects.

Study of the dynamics of stars and gas in the nuclear regions of
nearby galaxies has led to two remarkable discoveries: (i) the
overwhelming majority of spheroidal galaxies in the nearby universe
contain massive black holes and (ii) there is a correlation between
the black hole mass and the velocity dispersion of the stars in the
spheroid (Magorrian et al. 1998, Ferrarese \& Merritt 2000, Gebhardt
et al. 2000, Tremaine et al. 2002). Assuming that these relations also
apply in the high redshift universe (Shields et al. 2002), implies
that the highest redshift QSOs are associated with massive galaxies
($> 10^{11}~\rm{M_{\odot}}$).  Furthermore, strong metal emission
lines in these QSOs suggests that their environments have been quickly
enriched, possibly through starburst activity (Fan et al. 2001).

These fundamental results allow us to investigate the nature of these
earliest objects, and in particular to probe the relationship between
massive black hole and spheroidal galaxy formation. To address this
and other related questions, and in general, to study the cosmic
evolution of the radio-to-optical spectra of QSOs, we have undertaken
an extensive observational program from radio through millimeter
wavelengths of high redshift QSOs which includes: searches for
emission from warm dust at millimeter wavelengths from a large sample
of {z~$>$~1.8} QSOs (Carilli et al. 2001a, Omont et al. 2001, 2003),
high resolution imaging at cm wavelengths of the non-thermal radio
continuum emission from these sources (Carilli et al. 2001a, b), and
observations of the CO line emission from selected sources with large
infrared luminosities (Carilli, Menten \& Yun 1999; Carilli et
al. 1999, 2002a, b;  Cox et al. 2002; Beelen et al. in prep.).

Millimeter-continuum observations of high redshift QSO's show that
about 30$\%$ of the sources are detected in surveys with flux density
limits of 1 to 2 mJy at 250 GHz (Omont et al. 2001; Carilli et
al. 2001b; Isaak et al. 2002).  Multifrequency studies of the
rest-frame radio through far IR (FIR) spectral energy distributions
(SEDs) of these sources reveal that the mm emission is thermal
emission from warm dust (Benford et al. 1999), and in many of the
sources the cm-to-mm flux density ratios are consistent with the
radio-to-FIR correlation for star forming galaxies (Carilli et
al. 2001b).  Furthermore, searches for CO line emission from FIR
luminous QSOs have resulted in the detection of molecular line
emission indicating the presence of large gas reservoirs ($\sim
10^{11}$M$_{\odot}$; Cox et al. 2002). These detections lead some
authors to conclude that active star formation is inevitable (Omont et
al. 2001) in these FIR luminous, CO-rich quasars.  Andreani et
al. (2002) reached a similar conclusion for lower redshift QSOs,
although  for both low and high redshift samples
the case for star formation is by no means proven.

In this paper we extend our radio study of distant QSOs to the highest
redshifts ($z > 5$). We present observations at 1.4, 5, and 250
GHz. These observations are an order of magnitude, or more, more
sensitive than survey observations, such as FIRST and NVSS (Becker,
White, \& Helfand 1995; Condon et al. 1998). We assume a concordance
cosmology ($\Omega_m ~=0.3,~\Omega _{\Lambda}~ =~ 0.7$) and
H$_0$~=~65~km sec$^{-1}$~Mpc$^{-1}$ throughout.

\section{Observations}

We observed with the VLA all $z > 5$ QSOs known at the time of
observation, except J2245+0024 (Sharp, McMahon, Irwin \& Hodgkin
2001), which was outside our LST range. The sources are listed in
Table 1. Most of the sources come from the SDSS, and hence are
luminous QSOs ($\rm M_B < -26$), with the exception of J0301+0020,
which is a low luminosity QSO discovered by Stern et al. (2000b).

VLA observations at 1.4~GHz were made in the A configuration
(max. baseline = 30 km), with a total bandwidth of 100~MHz and two
polarizations. Each source was observed for about two hours.  In
addition, J0756+4104 was observed for 2 hours at 1.4~GHz in B
configuration (max. baseline = 10 km).  Standard phase and amplitude
calibration was applied, and all sources were self-calibrated using
field sources. The absolute flux density scale was set with
observations of either 3C48 or 3C286.

The final images were generated using the wide field imaging (Cotton
1999; Bridle \& Schwab 1999) and deconvolution capabilities of the
AIPS task IMAGR.  The theoretical rms noise ($\sigma$) value
corresponding to 2 hours of observing in continuum mode at 1.4~GHz is
{16~$\mu$Jy}, and in most of the maps presented here this sensitivity
is roughly achieved. The noise level in one source (J1204-0021) was
higher (35~$\mu$Jy), possibly due to some low level terrestrial
interference. We also include a $2\%$ uncertainty in absolute flux
density scale determination. The Gaussian restoring CLEAN beam Full
Width at Half Maximum (FWHM) was typically $\sim 1\farcs5$~ for the A
configuration observations.
 
Two of the sources, J0836+0054 and J0913+5919, were observed at 5~GHz
on August 17, 2002 in B configuration for about 20 minutes, achieving
an rms sensitivity of about 60$\mu$Jy\footnote{D. Rusin and
collaborators also observed J0836+0054 on May 5 at 5~GHz. Their flux
densities are equal to ours within the errors.}. The Gaussian
restoring CLEAN beam (FWHM) was $\sim1\farcs5~$, matching the
resolution of the 1.4GHz observations.

Observations at 250~GHz were made using MAMBO (Kreysa et al. 1999) at
the IRAM 30m telescope on Pico Veleta in Spain, during the winter of
2001-2002 within dynamically scheduled, pooled observations, except
for J0301+0020, which was reported on by Bertoldi \& Cox (2002).
MAMBO is a 37 element bolometer array with an effective central
frequency of 250 GHz for thermal sources.  The beam for the feed horn
of each bolometer is matched to the telescope beam of
10\farcs6. Observations were made in standard on-off mode, with 2 Hz
chopping of the secondary by 50$''$ in azimuth.  The data were reduced
using the MOPSI software package (Zylka 1998).  Pointing was monitored
every hour, and was found to be repeatable to within 2$''$. The sky
opacity was monitored every hour. Zenith optical depths were lower
than 0.3 at all times. Gain calibration was performed using
observations of Mars, Uranus, and Ceres.  We estimate a 20$\%$
uncertainty in absolute flux density calibration based on these
observations.

The target sources were centered on the central bolometer in the
array, and the temporally correlated variations of the sky signal (sky
noise) detected in the surrounding six bolometers were subtracted from
the central bolometer signal. The total sky + source observing time
for each source was about 1 hour, leading to rms sensitivities of 0.5
to 1.0 mJy, depending on the weather.

\section{Results}

The results from the 1.4 and 250~GHz observations are listed in Table
1. The abbreviated source names (column 1), the redshifts (column 2),
the absolute blue magnitude ($\rm M_{B}$) obtained by assuming a flat
$\Lambda$ dominated cosmology (column 3), and the optical position
(columns 4 and 5) are compiled from the optical discovery papers.
Columns 6 and 7 give the 1.4 and 250 GHz flux densities with 1$\sigma$
error-bars, and 3$\sigma$ upper limits are given for non-detections.
 
The 1.4 GHz images of these sources are shown in Figure 1. The
positional uncertainty for the radio observations is given by: $\sigma
_{\theta} \sim {{FWHM}\over{SNR}}$
(Fomalont 1999), where FWHM corresponds to that of the Gaussian
restoring beam, and SNR=Signal-to-Noise ratio of the detection. For a
3$\sigma$ detection this corresponds to 0\farcs5~ for most of our
sources. To this must be added the typical astrometric
uncertainty 0\farcs1~ (Pier et al. 2003) of the optical data, and the
uncertainty in the relationship between radio and optical reference
frame which is about 0\farcs 25 (Deutsch 1999). Thus, we only
consider emission within 0\farcs6~ of the optical position to be
associated with the QSO.

Fomalont et al. (2003) show that the sub-mJy source counts follow the
relation: $\rm N(>S_{1.4}) = 0.026 \times S_{1.4}^{-1.1}$
arcmin$^{-2}$, with 1.4 GHz flux density, S$_{1.4}$, in mJy.  Hence,
within 0\farcs6~ of a given source we expect $1.5\times 10^{-4}$
sources with $\rm S_{1.4} \ge 70\mu$Jy by chance.  Blain et al. (2002)
show that there are about 2000 sources deg$^{-2}$ with
S$_{250} > 1$ mJy. At this flux density level we then expect 0.01
sources by chance within the  beam of the 30m telescope 
(ie. within $5\arcsec$ of the target source).

\vskip 0.2in

\centerline{\bf Notes on individual sources}

{\bf J0231--0728 (z=5.41) } There is a 3$\sigma$ unresolved source at
1.4 GHz situated 1\farcs0~ away from the optical position of the
QSO. This is farther than the 0\farcs6~ positional accuracy of our
measurements so J0231-0728 is considered a non-detection at 1.4GHz.
This source is also not detected at 250 GHz.

{\bf J0301+0020 (z=5.50) } A radio continuum source with $\rm S_{1.4}
= 73 \pm 17\mu$Jy is detected within 0\farcs5~ of the optical QSO
position. The radio source is not resolved, and Gaussian fitting sets
an upper limit to its size at 1.4~GHz of 1\farcs6~. Bertoldi \& Cox
(2002) detect this QSO at 250~GHz with a flux density of $0.87\pm
0.20$ mJy.

{\bf J0756+4104 (z=5.09)} The combined A and B configuration
observations show a 1.4 GHz source with $\rm S_{1.4} = 65\pm 17 \mu$Jy
within $0\farcs2$ of the optical QSO position. Gaussian fitting to the
radio emission sets an upper limit to its size of 2\farcs3~. This QSO
is also detected with MAMBO with $\rm S_{250} = 5.5 \pm 0.5$ mJy.

{\bf{J0836+0054 (z=5.82) }} This source is clearly detected at 1.4~GHz
with a flux density of $1.75 \pm 0.04$ mJy.  Gaussian fitting sets an
upper limit to the source size of $0\farcs65$~ at 1.4 GHz.  J0836+0054
is also detected at 5~GHz with a flux density of $\rm S_{5} = 580 \pm
57 \mu$Jy.  The implied radio spectral index is --0.8. A second radio
source with S$_{1.4} = 0.44$mJy is detected 10\arcsec ~south of the
QSO. Deep optical and near-IR imaging of this field suggests that this
second source is associated with a lower redshift galaxy, and
unrelated to the QSO (Rusin et al. in prep). J0836+0054 is not
detected at 250~GHz with a $3\sigma$ upper limit of 2.9~mJy.

J0836+0054 was also detected in the FIRST radio survey (Becker, White
\& Helfand 1995) at 1.4~GHz with a flux density of 1.1$\pm$0.15 mJy.
The difference in flux density between the FIRST measurement and our
more recent measurement is significant at the 4$\sigma$ level, such
that the radio source appears to be variable on yearly time-scales.

{\bf{J0913+5919 (z=5.11)}} This source is clearly detected at 1.4~GHz
with $\rm S_{1.4} = 18.95\pm 0.4$ mJy.  Gaussian fitting to the radio
emission sets an upper limit to the source size of $0\farcs2$~ at 1.4
GHz. J0913+5919 is also clearly detected at 5 GHz with $\rm S_{5} =
8.1 \pm 0.2$ mJy, implying a spectral index for this source of --0.7.
This source is not detected at 250~GHz with a $3\sigma$ upper limit of
2.8~mJy.

J0913+5919 was also detected in the NVSS radio survey (Condon et
al. 1998) with a total flux density of 18.5$\pm$0.5~mJy, implying
that the source is not variable on yearly time-scales.

{\bf J1030+0524 (z=6.28)} This source was the highest redshift QSO
published at the time of observation. The source is not detected at
1.4 or 250 GHz.  However, four fairly bright sources ($\rm S_{1.4} >
200 \mu$Jy) are detected at 1.4 GHz in a 2\arcmin ~field centered on the
QSO (Figure 2), including an edge-brightened double radio source
('Fanaroff-Riley Class II'; Fanaroff \& Riley 1974) with an extent of
about 1\arcmin ~and a total flux density of 29 mJy.

{\bf J1044-0125 (z=5.73)} This source is not detected at 1.4~GHz with
a 3$\sigma$ upper limit of S$_{1.4} < 79 \mu$Jy. We did not observe
this source with MAMBO.  Iwata et al. (2001) report a flux density at
350 GHz for J1044--0125 of $6.2\pm 2.0$ mJy.  Assuming a thermal IR
spectrum typical for Ultraluminous IR galaxies leads to an expected
250 GHz flux density of $3.4\pm 1.1$ mJy for a source at $z = 5.73$.
Near-IR spectroscopic observations of this source reveal a prominent
CIV absorption feature whose shape suggests that this is a Broad
Absorption Line QSO (Maiolino et al. 2001; Goodrich et al. 2001).

\section{Analysis}

\subsection{Fraction of radio loud objects}

Questions regarding both the bi-modality of the radio luminosity
function of QSOs, and its evolution, have been investigated by numerous
workers (e.g. Peacock, Miller \& Longair 1986; Miller,
Peacock, \& Mead 1990; Schmidt, van Gorkom, Schneider, \& Gunn 1995; Stern et al. 2000;
Lacy et. al 2001; Ivezic et. al. 2002).  Two definitions are generally
used to demarcate radio-quiet and radio-loud QSOs. One criterion is
based on the radio-optical ratio, R$_{\rm{ro}}$, of the specific fluxes at
rest-frame 5GHz and 4400\AA~ (Kellerman et al. 1989) where typical
radio--loud sources have R$_{\rm{ro}}$ in the range 10--1000, while
radio quiet sources have R$_{\rm{ro}} < 1$.
Peacock, Miller \& Longair (1986) point out that R$_{\rm{ro}}$ can 
be used as a discriminating parameter only if the radio
and optical luminosities are linearly correlated, which does not seem
to be the case (Stocke et al. 1992). The second definition divides the
sources at the 1.4 GHz luminosity density of ~$3\times 10^{25}$ W
Hz$^{-1}$ (Gregg et al. 1996).  Ivezic et al. (2002) find that for
optically selected QSOs the two definitions are consistent as a
consequence of selection effects in flux limited samples,
 so for the remainder of this paper we will use
the Gregg et al. (1996) definition to classify a source as radio loud.

Having the radio spectral indices of the two bright radio sources,
J0836+0054 and J0913+5919 ($-0.8$ and $-0.7$, respectively), allows us
to estimate their luminosity densities at a rest frame frequency of
1.4~GHz. For J0836+0054 the value is $5.0\times 10^{26}$ W Hz$^{-1}$,
while that for J0913+5919 is $4.9\times 10^{27}$~ W Hz$^{-1}$. 
Both these sources are radio-loud by any definition.

The radio spectral indices for J0756+4104 and J0301+0020 are unknown,
so we calculate luminosity densities using the spectral index of
$-0.5$ adopted in previous studies (eg. Stern et al. 2000, Ivezic et
al. 2002).  The 1.4~GHz rest frame luminosities for these two sources
are $1.2\times 10^{25}$ and $1.6\times 10^{25}$ W Hz$^{-1}$
respectively, placing them below the radio-loud demarcation.

\subsection{The radio-FIR correlation}

Three sources (J0756+4104, J0301+0020 and J1044-0125) are detected at
(sub)mm wavelengths at levels at least ten times above their 1.4 GHz
flux densities, implying rapidly rising spectra from cm-to-mm
wavelengths.  Note that an observing frequency of 250 GHz corresponds
to a rest frequency of 1600 GHz (= 188 $\mu$m) for a source at $z =
5.5$, such that the MAMBO observations sample the rest frame far IR
part of the spectra. Benford et al.  (1999) have performed
multiwavelength observations of a number of mm-loud high redshift
QSOs, and in every case they find that the rest frame far IR SEDs are
consistent with a gray body spectrum characteristic of thermal
emission from warm dust.  In the analysis below we will assume that
the rapidly rising cm-to-mm spectra of the three mm-loud sources in
our sample imply thermal emission from warm dust. However, we cannot
preclude more exotic explanations for the rapidly rising spectra,
such as synchrotron self absorption at rest frame far IR wavelengths.

One possible method for addressing the question of dust heating is
using the fact that star-forming galaxies at low redshift follow a
very tight linear relation between radio continuum and FIR luminosity
(Condon \& Yin 1990; Condon 1992; Crawford, Marr, Partridge \& Strauss
1996; Miller \& Owen 2001; Yun, Reddy, \& Condon 2001). This
correlation holds over four orders of magnitude in luminosity with
only a factor of two scatter around linearity for galaxy samples
selected in the optical, IR, and radio. While a general correlation is
expected since the synchrotron radiation at centimeter wavelengths,
and thermal dust emission at IR wavelengths, both relate to massive
star formation, the tightness and linearity of the correlation remain
puzzling. A further uncertainty in using the FIR-radio correlation, to
assess the significance of star-formation in heating the dust, is that
lower luminosity radio quiet QSOs at lower redshift also follow the
standard radio--FIR correlation for star forming galaxies (Sopp \&
Alexander 1991). Whether the sources in the Sopp \& Alexander sample
also host active star formation remains unknown.  Also, there is the
question of whether the radio-FIR correlation holds in star forming
galaxies at high redshift (Carilli \& Yun 1999; 2000; Yun \& Carilli
2002; Garrett 2001). Given these uncertainties, we feel that comparing
centimeter and millimeter continuum luminosities is simply a
consistency check for star formation, but not conclusive evidence for
it.

Figure 3 shows the relationship between redshift and the 250-to-1.4
GHz spectral index ($\alpha^{250}_{1.4}$) for a star-forming galaxy
taken from the study of Carilli \& Yun (2000).  This model consists of
the mean SED (plus scatter) of 17 low-redshift star-forming galaxies.
The far IR part of this model SED corresponds roughly to a modified
blackbody spectrum with a dust temperature of 50K and emissivity index
of 1.5, and the centimeter part of the spectrum is constrained to
follow the radio-to-FIR correlation for star forming galaxies.  A
point located below the curve (plus scatter) on this diagram indicates
a source that is radio loud relative to a star-forming galaxy.  In
such an object it is likely that the radio emission is due to the AGN
and not star formation (Yun et al. 2001).

For J0301+0020 we find $\alpha^{250}_{1.4} = 0.5
\pm 0.1$, well below the region defined by star forming galaxies. For
J0756+4104 we find $\alpha^{250}_{1.4} = 0.86 \pm 0.08$, which is
consistent (within the errors) with the low end of the range for star
forming galaxies. For J1044-0125 we extrapolate to 250 GHz from the
measured 350 GHz flux density (section 3). This source was not
detected at 1.4 GHz, implying $\alpha^{250} _{1.4} > 0.7$, which is at
least consistent with star formation. A factor two deeper radio
observations are required to test whether this source follows the
radio-FIR relationship defined by low redshift galaxies.

\subsection{Inverse Compton emission from J1306+0356?}

X-ray observations of the source J1306+0356 (Brandt et al. 2002)
suggest a possible jet-like feature 23\arcsec~ from the QSO position
(Schwartz 2002).  Schwartz (2002) argues that the emission may be due
to inverse Compton (IC) scattering of the cosmic background by
relativistic electrons in a jet emanating from the QSO. 
These same relativistic electrons would emit radio synchrotron
radiation in the presence of a magnetic field, hence we have
searched for radio emission from the location of the jet-like X-ray
feature.  

We have not detected radio emission from the possible X-ray
jet in J1306+0356 to a 3$\sigma$ limit of 150 $\mu$Jy at 1.4 GHz
(after convolving to the $5''\times2''$ resolution corresponding to
the box containing the X-ray 'jet'). The flux density of the X-ray
feature at 1 keV ($2.4\times 10^{17}$ Hz) is $8.3\times 10^{-4}\mu$Jy,
assuming a spectral index of $-1$.

A long standing and well documented technique for deriving magnetic
fields in extragalactic radio sources is by comparing the radio
synchrotron and X-ray IC flux densities (Harris \& Grindlay 1979). The
constraint on the magnetic field strength comes from the fact that the
IC X-ray emissivity is a function of the relativistic electron density
and the energy density in the dominant ambient photon field
(presumably the microwave background), and the synchrotron radio
emissivity is a function of the relativistic electron density and the
magnetic energy density.  A recent simple parameterization of this
calculation can be found in equation 4 in Carilli \& Taylor (2002).
In the case of the jet-like X-ray feature in J1306+0356, only a radio
upper limit is available, such that we can only derive an upper limit
to the magnetic field strength.  Using equation 4 from Carilli \&
Taylor (2002), and the X-ray flux density and radio upper limits given
above, we derive an upper limit of 3$\mu$G to the magnetic field in
the jet-like feature in J1306+0356. Note that this limit assumes the
X-ray emission is IC, and that the relativistic electrons have a
power-law energy distribution of index --3 over a wide range in energy
(see Section 5.3 for more details).

\section{Discussion}

\subsection{Radio loud QSOs at high redshift}

There has been considerable debate in the literature about the
redshift evolution of the radio-loud fraction of QSOs (e.g. Visnovsky et al. 1992; 
Schneider, van Gorkom, Schmidt \& Gunn 1992; Schmidt, van Gorkom, Schneider, \& Gunn 1995;
 Hooper et al. 1995; Goldschmidt et
al. 1999; Stern et al. 2000). Most recently Ivezic
et al. (2002) have investigated this question using large samples of
QSOs out to $z \sim 2.2$ from the SDSS. They conclude that about 8$\%$
of QSOs are radio-loud, independent of redshift.  Two out of the ten
QSOs at $z > 5$ observed in our study are radio-loud, which is roughly
consistent (given the small number statistics) with the fraction seen
at lower redshift, eg. the Poisson probability of seeing 2 radio--loud
quasars when the expectation value is 0.8 is 14$\%$. As emphasized by
Ivezic et al (2002), larger QSO samples at high redshift are required
to separate effects related to redshift-luminosity biases in flux
limited samples.

The radio loud sources J0836+0054 and J0913+5919 are compact ($<
1$\arcsec~) and have steep spectra (--0.8 and --0.7), classifying them
as Compact Steep Spectrum (CSS) objects.  CSS objects are a mixed-bag
of source types, ranging from steep spectrum core-jets (size $\le$ few
pc), to small (size $\sim 1$ kpc), double-lobed radio galaxies,
ie. compact symmetric objects (CSOs; Tzioumis et al. 2002). The CSOs
are particularly intriguing since they are thought to be very young
radio sources, and their ages can be measured through VLBI
observations of the proper motions of the radio hot spots, with
typical measured ages between 10$^3$ and 10$^4$ years (Polatidis \&
Conway 2002).  Recent VLBI imaging of J0836+0054 by Frey et al (2003)
at 1.6 GHz shows a marginally resolved source at 10 mas resolution,
suggesting that this source is a steep spectrum core-jet, and not a
CSO. Similar VLBI imaging has not been performed for J0913+5919, but
the lack of variability of this source (section 3) makes it a prime
candidate for a high redshift CSO.

In either case, the fact that the radio-loud fraction of QSOs appears
to be relatively constant out to very high redshift, and that the two
radio-loud QSOs at $z > 5$ discovered thus far have steep radio
spectra, is encouraging from the perspective of studying the neutral
intergalactic medium (IGM) during the epoch of reionization through
observation of HI 21cm absorption (the '21cm Forest'). Calculations by
Carilli, Gnedin, \& Owen (2002), and Furlanetto \& Loeb (2002) show
that the next generation low frequency, large area radio telescopes,
such as the LOFAR and the Square Kilometer Array, will be able to
study the neutral IGM beyond the epoch of reionization (EOR) in
\ion{H}{1} 21cm absorption toward discrete radio sources as faint as a
few mJy at frequencies below 200 MHz. Both J0836+0054 and J0913+5919
would be easily adequate for such studies if they were placed beyond
the EOR.

\subsection{Thermal emission from warm dust and star formation}

Three of the ten sources in our study (J0301+0020, J0756+4104, and
J1044-0125) have been detected at (sub)mm wavelengths with flux
densities much larger than their 1.4 GHz flux densities.  Again, such
steeply rising spectra provide strong evidence that the (sub)mm
emission is thermal emission from warm dust.  

We have considered these sources in the context of the radio-to-FIR
correlation for star forming galaxies, as quantified in the 1.4-to-250
GHz spectral index, $\alpha^{250} _{1.4}$ (Figure 3).  For J0301+0020
the $\alpha^{250}_{1.4}$ value is clearly below that expected for a
star forming galaxy.  In this case it seems likely that the radio
emission relates to the AGN activity. For J0756+4104 the
$\alpha^{250}_{1.4}$ value is within the range defined by star forming
galaxies, while for J1044--0125 the lower limit to
$\alpha^{250}_{1.4}$ is at least consistent with a star forming
galaxy.

Given the interest in co-eval massive black hole--spheroidal galaxy
formation, it is instructive to consider what the properties of these
sources would be if the 250 GHz emission were a result of dust heated
by a starburst. Using the relations between 250 GHz flux density, FIR
luminosity, and star formation rates, in Omont et al. (2003) based on
typical spectra of low redshift ultra-luminous infrared galaxies
(i.e. Arp 220) leads to: $L_{\rm FIR} = 2.8\times 10^{12}$ L$_\odot$
for J0301+0020, $1.4\times 10^{13}$ L$_\odot$ for J0756+4104, and
$7\times 10^{12}$ L$_\odot$ for J1044--0125\footnote{Shioya et
al. (2003) have found a faint optical galaxy located within 1\farcs9
of the quasar.  They suggest that the QSO may be gravitationally
magnified by this galaxy by a factor of about two. We have not
corrected for this magnification in the calculations above.}, and
massive ($>5$ M$_\odot$) star formation rates of 200, 1000, and 500
M$_\odot$ year$^{-1}$, respectively.  At these extreme star-formation
rates most of the stars in a large spheroidal galaxy could form in a
dynamical time of $10^8$~years. However, it should again be stressed
that these rates assume the dust is heated by star formation, as
opposed to being heated by the AGN itself.
     
\subsection{X-ray-loud, radio-quiet jets at high redshift?}

We have not detected radio emission from the reported X-ray 'jet' in
J1306+0356 (Schwartz 2002). The implied upper limit to the magnetic
field (3$\mu$G) is more than an order of magnitude below typical
magnetic field values in powerful radio jets. Perhaps the simplest
conclusion from these results is that the X-ray feature is not IC
emission from a jet emanating from J1306+0356.

It is possible that such a jet could be Inverse Compton X-ray-loud and
still be radio-quiet, even for strong magnetic fields, due to the
different radiative lifetimes of the particles involved. The spectral
peak of the CMB behaves as $1.6\times10^{11}$(1+z) Hz, such that
observations at 1 keV are sensitive to electrons with Lorentz factors,
$\gamma_e \sim 1000$, independent of redshift. The radiative lifetime
of such electrons at $z \sim 6$ is $\sim 1.0\times10^7$ years. For
comparison, observations of a $z=6$ jet at 1.4 GHz probe electrons
with $\gamma_e \sim 7000$, assuming a 50$\mu$G magnetic field,
corresponding to a radiative lifetime $\sim 1.3\times10^6$ years. If
particle acceleration ceased between 10$^6$ and 10$^7$ years ago, then
the exponential cut-off at high energies in the relativistic particle
population might lead to the situation observed, ie. an X-ray-loud but
radio-quiet jet. More sensitive X-ray observations, and lower
frequency radio observations, are required to test this interesting
possibility.

\subsection{Over-densities of radio sources toward the highest redshift
QSOs}

For the highest redshift QSO in our sample, J1030+0524 at $z=6.28$
(Figure~2), we find four fairly bright radio sources ($\rm S_{1.4} >
200\mu$Jy) within 1\arcmin~ of the QSO, one of which is an arcmin-scale
powerful double (FRII) radio source.  Note that for a 2\arcmin~ field
one expects only 0.5 sources by chance with S$_{1.4} > 200\mu$Jy.  It
seems unlikely that the sources are at the redshift of the QSO, since
an arcmin-scale double radio galaxy has never been detected beyond $z
\sim 3$ (Carilli et al. 1997).  Interestingly, the highest redshift
QSO known, J1148+5251 at $z = 6.4$ (Fan et al. 2003), also has two
bright radio sources (8 and 70 mJy at 1.4 GHz) within 1\arcmin~ of
the QSO position (Bertoldi et al. 2003).

A possible explanation for the excess radio source density in the
fields of the two highest redshift QSOs is that the radio sources are
in a group or cluster that happens to lie along the line-of-sight.
If this is the case, then gravitational lensing by the cluster could
magnify
the QSO emission by a factor 2 or so without giving rise to
arcsecond-scale distortions in the optical images of the QSOs.

\vskip 0.2in

The National Radio Astronomy Observatory is
operated by Associated Universities, Inc., under a cooperative
agreement with the National Science Foundation.
M.A.S. acknowledges support of NSF grant AST-0071091.

 \begin{deluxetable}{lcccccc}
  \tablecolumns{7}
  \tablecaption{Properties of Observed QSOs}
  \tablehead{
   \colhead{} 
    &\colhead{}
    &\colhead{} 
    &\multicolumn{2}{c}{Optical Position (J2000)}
    &\colhead{S$_{1.4}$}
    &\colhead{S$_{250}$}\\
     \colhead{QSO } 
    &\colhead{$z$}
    &\colhead{$\rm M_{B}$} 
    &\colhead{RA ($h~m~d$)} &\colhead{DEC (\arcdeg ~ \arcmin ~\arcsec~)}
    &\colhead{[$\mu$Jy]}
    &\colhead{[mJy]}\\
    }
  \startdata
  J0231--0729\tablenotemark{a}&5.41 &-27.37 & 02 31 37.65 & -07 28 54.5
   & $<50$ & $<3.5$\\
  J0301+0020\tablenotemark{b} &5.50 &-24.00 & 03 01 17.01 &  00 20
  26.0 & 73~$\pm $~18& 0.87~$\pm $~0.2 \\ 
  J0756+4104\tablenotemark{a}  & 5.09 &-26.50 & 07 56 18.14 &  41 04
  08.6 & 65 ~$\pm $~17 & 5.5$\pm$0.5\\ 
  J0836+0054\tablenotemark{c}  & 5.82 &-28.10 & 08 36 43.85 &  00 54
  53.3 &  1750~$\pm$~40 & $<2.9$\\ 
  J0913+5919\tablenotemark{a}  &5.11 &-26.20 & 09 13 16.56 &  59 19
  21.5 & 18950~$\pm$~380 & $<2.8$\\ 
  J1030+0524\tablenotemark{c}  &6.28 &-27.37 & 10 30 27.10 &  05 24 
  55.0 &$<61$ & $<3.4$ \\
  J1044--0125\tablenotemark{d}& 5.73 &-27.63 & 10 44 33.04 & -01 21
  49.6 & $<79$ & 3.4$\pm$1.1\tablenotemark{e} \\ 
  J1204--0021\tablenotemark{f}& 5.03 &-27.64 & 12 04 41.73& -00 21
  49.6 &$<87$ & NA \\ 
  J1208+0010\tablenotemark{g}  & 5.27 &-26.30 & 12 08 23.82 & 00 10
  27.7 &$<60$ & $<3.1$ \\ 
  J1306+0356\tablenotemark{c}  & 5.99 &-27.41 & 13 06 08.26 &  03 56
  26.3 & $<53$ & $<3.1$ \\ 
  \enddata	
  \tablenotetext{a}{Anderson et al. 2001}
  \tablenotetext{b}{Stern et al. 2000b}
  \tablenotetext{c}{Fan et al. 2001}
  \tablenotetext{d}{Djorgovski et al. 2001; Fan et al. 2000b}
  \tablenotetext{e}{Extrapolated from 350 GHz (section 3)}
  \tablenotetext{f}{Fan et al. 2000a}
  \tablenotetext{g}{Zheng et al. 2000}
\end{deluxetable}

\clearpage

\centerline{Figure Captions}
{\bf Figure 1:} Images at 1.4~GHz of the ten high redshift quasars discussed in 
this paper. The FWHM of
the Gaussian restoring beams are shown in the insets in all frames. 
Contour levels (solid lines) are a geometric
progression in the square root of two starting at 2$\sigma$, with
$\sigma$ listed below
($\sigma$ corresponds to the measured rms on the image). Two negative
contours (dashed) are included.  The central cross in each image
marks the optical QSO location.  \\
 Fig1a = J0231-0728, $\sigma$ = 16.6 $\mu$Jy beam$^{-1}$; 
 Fig1b = RD J0301+0020, $\sigma$ = 17.2 $\mu$Jy beam$^{-1}$;
 Fig1c = J0756+4104, $\sigma$ = 17 $\mu$Jy beam$^{-1}$;
 Fig1d = J0836+0054, $\sigma$ = 21.6 $\mu$Jy beam$^{-1}$;
 Fig1e=  J0913+5919, $\sigma$ = 14 $\mu$Jy beam$^{-1}$;
 Fig1f=  J1030+0524, $\sigma$ = 20.3 $\mu$Jy beam$^{-1}$;
 Fig1g=  J1044-0125, $\sigma$ = 26.5 $\mu$Jy beam$^{-1}$;
 Fig1h=  J1204-0021, $\sigma$ = 29 $\mu$Jy beam$^{-1}$;
 Fig1i=  J1208+0010, $\sigma$ = 20 $\mu$Jy beam$^{-1}$;
 Fig1j=  J1306+0356, $\sigma$ = 17.8 $\mu$Jy beam$^{-1}$;

{\bf Figure 2:} A wider field image of J1030+0524 at 1.4~GHz.  The
FWHM of the Gaussian restoring beam is $3\arcsec~ \times 3\arcsec~$.
Contour levels are a geometric progression in the square root two
starting at 0.13 mJy beam$^{-1}$. Three negative contours (dashed) are
included. The cross shows the position of the optical QSO.

{\bf Figure 3:} The relationship between redshift and the observed
spectral index between 250 and 1.4~GHz for star-forming galaxies
(solid curve), as derived from the models presented in Carilli \& Yun
(2000). The dashed lines show the rms scatter in the distribution. The
250--to--1.4 GHz spectral indexes for QSOs detected at 250~GHz are
shown with 1$\sigma$ error bars.



\clearpage\newpage
    
\begin{figure}
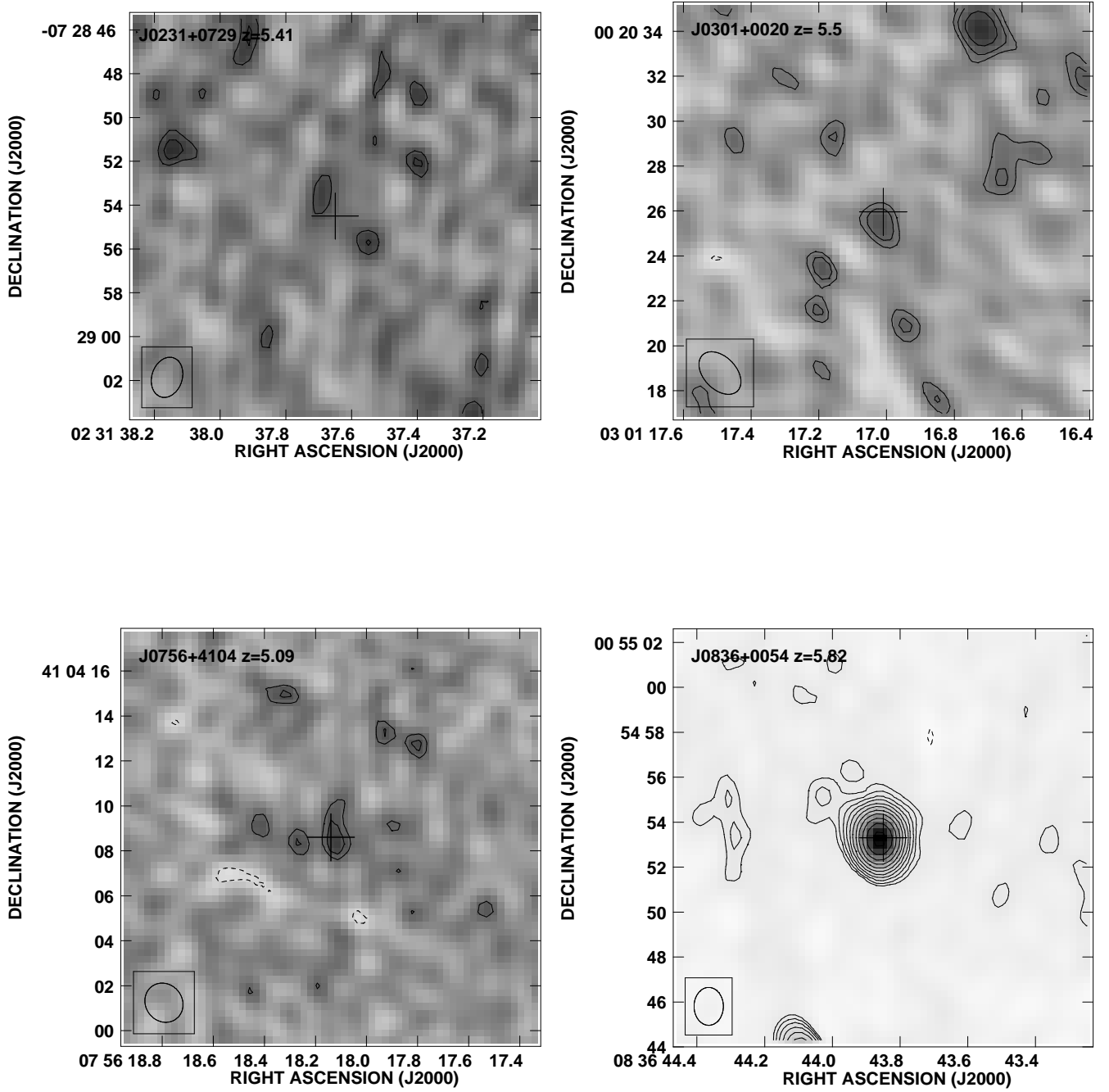

\figurenum{1}
\psfig{figure=PetricFig1a.ps,width=3.5in}
\vskip -4in
\hspace*{3.5in}
\psfig{figure=PetricFig1b.ps,width=3.5in}
\psfig{figure=PetricFig1c.ps,width=3.5in}
\vskip -4in
\hspace*{3.5in}
\psfig{figure=PetricFig1d.ps,width=3.5in}
\caption{Upper left: figure 1a, upper right: figure 1b, lower left: figure 1c,
lower right: figure 1d}
\end{figure}

\clearpage\newpage

\begin{figure}
\figurenum{1cont.}
\psfig{figure=PetricFig1e.ps,width=3.5in}
\vskip -4in
\hspace*{3.5in}
\psfig{figure=PetricFig1f.ps,width=3.5in}
\psfig{figure=PetricFig1g.ps,width=3.5in}
\vskip -4in
\hspace*{3.5in}
\psfig{figure=PetricFig1h.ps,width=3.5in}
\caption{Upper left: figure 1e, upper right: figure 1f, lower left figure 1g,
lower right figure 1h}
\end{figure}

\clearpage\newpage
\begin{figure}
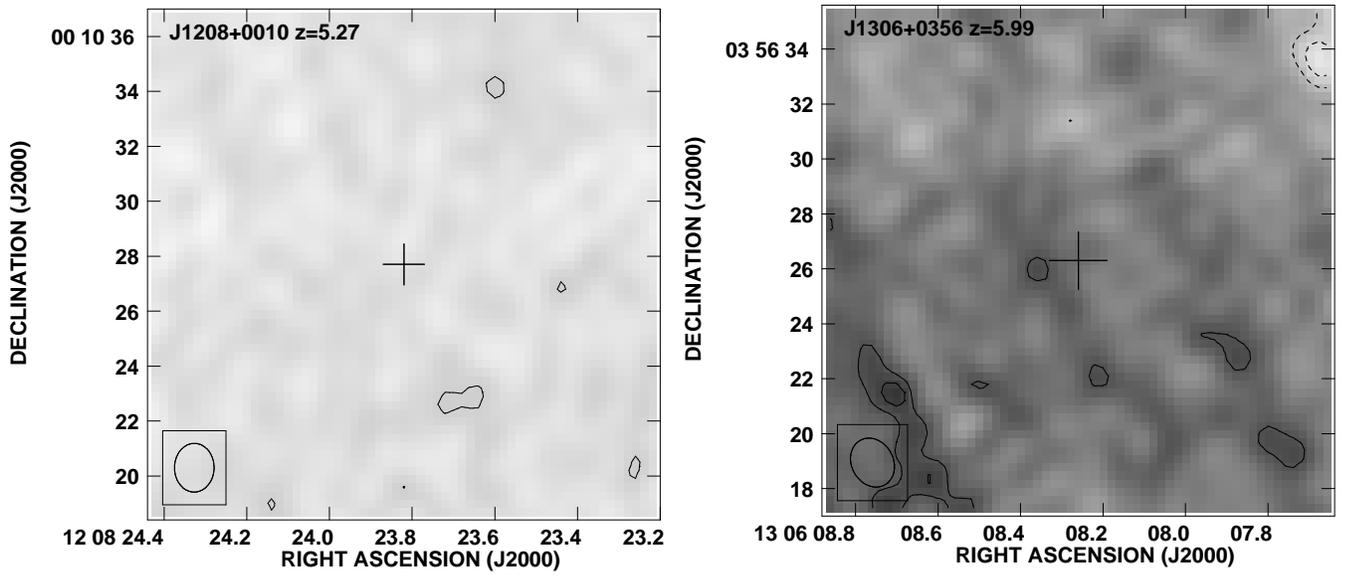

\figurenum{1cont.}
\psfig{figure=PetricFig1i.ps,width=3.5in}
\vskip -3.92in
\hspace*{3.5in}
\psfig{figure=PetricFig1j.ps,width=3.5in}
\caption{Left: figure 1i, Right: figure 1j}
\end{figure}

\clearpage\newpage

\begin{figure}
\figurenum{2}
\psfig{figure=PetricFig2.ps,width=7in}
\end{figure}

\clearpage\newpage

\begin{figure}
\figurenum{3}
\psfig{figure=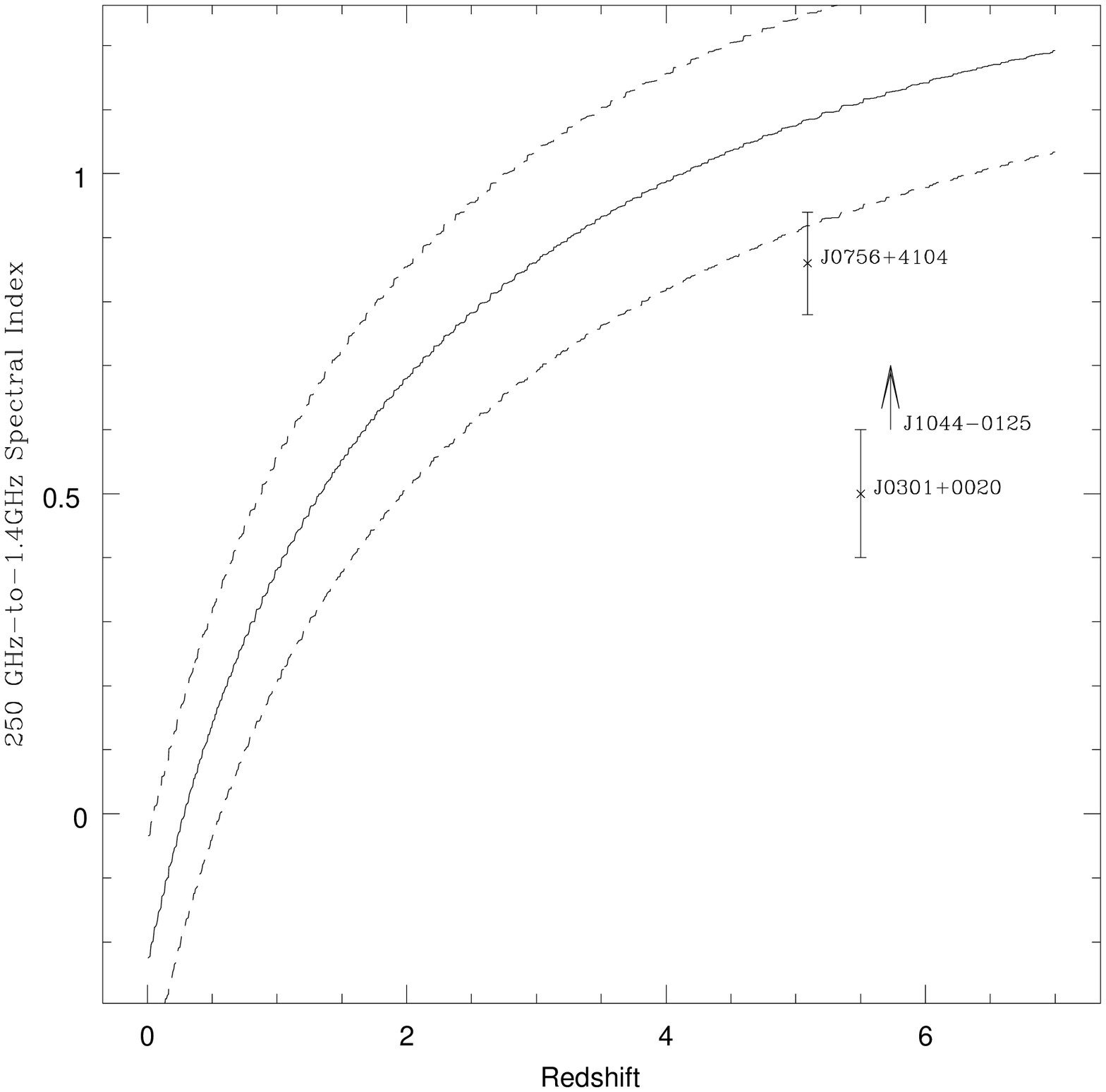,width=6in}
\end{figure}

\begin{references}
\reference{} Anderson, S. F., Fan, X., Richards, G.T. et al. 2001, AJ
122, 503 
\reference{} Andreani, P., Cristiani, S., Grazian, A., La
Franca, F., \& Goldschmidt, P. 2002, AJ, in press 
\reference{} Becker,
R. H., White, R.L., Helfand, D. J., 1995, ApJ, 450, 559 
\reference{} Becker, R. H., Fan, X., White, R. L., Strauss, M.A. et
al. 2001, AJ, 122, 2850 
\reference{} Beelen, A., 2003, in prep.
\reference{} Benford, D. J., Cox, P., Omont, A., Phillips, T.G., \& McMahon, R.G., 1999,
ApJ 518, 65
\reference{} Bertoldi, F., \& Cox, P., 2002, A\&A, 384L, 11
\reference{} Blain, A.W., Smail, I., Ivison, R.J., Kneib, J.-P., \&
Frayer, D.T., 2002, Phys. Rept. 369, 111 
\reference{} Brandt, W.N., Schneider, D.P., Fan, S., Strauss, M.A., et al. 2002, ApJ, 569, L5
\reference{} Bridle, A.H. \& Schwab, F.R. 1999, \ Bandwidth and
Time-Average Smearing, in \ Synthesis Imaging in Radio Astronomy II \
eds. Taylor, G.B., Carilli, C.L., \& Perley, R.A., Astronomical
Society of the Pacific Conference Series vol. 180, p. 371 
\reference{}Carilli, C.L., Bertoldi, F., Omont, A., Cox, P. et
al. 2001a, AJ, 122, 1679 
\reference{} Carilli, C.L., Bertoldi, F., Rupen, M.P., Fan, X.,
et al. 2001b, ApJ, 555, 625 
\reference{} Carilli, C.L., Cox, P.,
Bertoldi, F., Menten, K.M, et al. 2002b, ApJ, 575, 145 
\reference{}Carilli, C.L., Gnedin, N.Y. \& Owen, F. 2002, ApJ 577, 22 
\reference{}Carilli, C.L., Kohno, K., Kawabe, R., Ohta, K., et
al. 2002a, 123, AJ, 1838 
\reference{} Carilli, C.L., Rottgering, H., Miley, G., Pentericci, L.,
\& Harris, D. 1997, in {\sl The most distant radio galaxies},
eds. H. Rottgering, P. Best, \& M. Lehnert, (Amsterdam: Royal Dutch
Academy), p. 123  
\reference{} Carilli, C.L. \& Taylor, G.B. 2002, ARA\&A, 40, 319 
\reference{} Carilli, C.L. \& Yun, M.S., 2000, ApJ, 530, 618 
\reference{} Carilli, C.L. \& Yun, M.S., 1999, ApJ, 513, 13 
\reference{} Carilli, C.L., Menten, K.M., \& Yun, M.S.,1999, ApJ, 521,
L25 
\reference{} Condon, J.J., Cotton, W.D., Greisen, E. W., Yin, Q. F.,
et al. 1998, AJ, 115, 1693 
\reference{} Condon, J.J. 1992, ARA\&A, 30, 575 
\reference{} Condon, J.J. \& Yin, Z.F. 1990, ApJ 357, 97 
\reference{} Cotton, W.D., 1999,
in {\sl Synthesis Imaging in Radio Astronomy II}, ASP Vol. 180,
eds. Taylor, G.B., Carilli, C.L., \& Perley, R.A., (San Francisco:
Astronomical Society of the Pacific), p. 357 
\reference{} Cox, P., Omont, A., Djorgovski, S.G., Bertoldi, F., et al. 2002, A\&A, 387, 406
\reference{} Crawford,
T., Marr, J., Partridge, B., \& Strauss, M.A., 1996, ApJ, 460, 225
\reference{} Dale, D., Silbermann, N.A, Helou, G., Valjavec, E., et al. 2000, AJ, 120, 583
\reference{} Dale, D., Helou, G., Contursi, A., Silbermann, N., \& Sonali, K., 2001, ApJ, 549, 215
\reference{} Deutsch, E. W., 1999, AJ, 118, 1882 
\reference{} Djorgovski, S. G., Castro, S., Stern, D., Mahabal,
A. A. 2001, ApJ, 560, L5 
\reference{} Elvis, M., Wilkes, B.J., McDowell, J.C., Green,
R., et al. 1994, ApJS, 95, 1 
\reference{} Fan, X., Strauss, M.A., Schneider, P., Gunn, J.E., et al. 2000a, AJ 119, 1
\reference{} Fan, X., White, R.L., Davis, M., Becker, R. H., et al. 2000b, AJ, 120, 1167 
\reference{} Fan, X., Narayanan, V., Lupton, R. H., Strauss, M. A., et
al., 2001, AJ, 122, 2833 
\reference{} Fan, X., Strauss, M.A., Becker, R., Schneider, D. 
et al. 2003, AJ, in press
\reference{} Fanaroff, B.L. \& Riley, J.M. 1974, MNRAS, 167, 31
\reference{} Ferrarese, L., \& Merritt, D. 2000, ApJ, 539, L9
\reference{} Fomalont, E. B., 1999 in {\sl Synthesis Imaging in Radio
Astronomy II}, ASP Vol. 180, eds. Taylor, G.B., Carilli, C.L., \&
Perley, R.A., (San Francisco: Astronomical Society of the Pacific),
301 
\reference{} Fomalont, E.G., 2002 in prep.  
\reference{} Frey, S.,
Mosoni, L., Paragi, Z., \& Gurvits, L. MNRAS 2003, submitted
\reference{} Furlanetto, S.R. \& Loeb, A. 2002, ApJ, 579, 1
\reference{} Garrett, M.A. 2001, in {\sl The
Central Kiloparsec of Starbursts and AGN, ASP Vol. 249}, eds
J. H. Knapen, J. E.  Beckman, I. Shlosman, and T. J. Mahoney (San
Francisco: Astronomical Society of the Pacific), 652 
\reference{} Gebhardt, K., Kormendy, J., Ho, L., Bender, R., et
al. 2000, ApJ, 543, L5 
\reference{} Goldschmidt, P., Miller, L., La Franca, F., \& Christiani, S. 1992, MNRAS, 256, P65
\reference{} Goodrich, R.W., Campbell, R., Chaffee, F.H, Hill, G.M.,
et al. 2001, ApJ, 561, L23 
\reference{} Gregg, M. D., Becker, R. H., White, R.L.,
Helfand, D.J., et al. AJ, 112, 470
\reference{} Gunn, J.E. \& Peterson, B. A. 1965, ApJ, 142, 1633 
\reference{}Harris, D.E. \& Grindlay, J.E. 1979, MNRAS, 188, 25
\reference{} Hooper, E.J., Impey, C.D., Foltz, C.B., \& Hewett,
P.C. 1995, ApJ, 445, 62 
\reference{} Isaak, K. et al. 2002, MNRAS, 329, 149
\reference{} Ivezic, Z., Menou, K., Strauss, M., Knapp,
G.R., et al. 2002, AJ, 124, 2364 
\reference{} Iwata, I., Ohta, K.,
Nakanishi, K., Kohno, K. \& McMahon, R. 2001, PASJ, 53, 871
\reference{} Kreysa, E., Gemuend, H.P., Gromke, J., Haslam, C.G., et al. 1999, SPIE, 3357, 319 
\reference{} Magorrian, J., Tremaine, S., Richstone, D., Bender, R.,
et al. 1998, AJ, 115, 2285 
\reference{} Maiolino, R., Mannucci, F., Baffa, C., Gennari, S., \&
Oliva, E., 2001, A\&A, 372, L5 
\reference{} Miller, L., Peacock,
J. A., Mead, A. R. G. 1990, MNRAS, 244, 207 
\reference{} Omont, A.,
Cox, P., Bertoldi, F., McMahon, R.G., et al. 2001, A\&A, 374, 371
\reference{} Omont, A., Beelen, A., Bertoldi, F. et al. 2003, A\&A, in
press 
\reference{} Peacock, J. A., Miller, L., \& Longair, M.S., 1986
MNRAS 218, 265 
\reference{} Pentericci, L., Fan, X., Rix, H.-W., Strauss, M. A., et
al. 2002, AJ, 123, 2151 
\reference{} Pier, J.R., Munn, J.A., Hindsley,
R.B., Hennessy, G.S., et al. 2003, AJ in press, ASTROPH 0211375
\reference{} Polatidis, A.G. \& Conway, J.E. 2002, in {\sl The third
GPS/CSS workshop}, Pub. Astron. Soc. Australia, Vol 20, eds.
Tzioumis, T., de Vries, W., Snellen, I., Koekemoer, A. (ASA: Sydney)
\reference{} Schmidt, M., van Gorkom, J.H., Schneider, D.P., Gunn,
J. E. 1995, AJ 109, 473 
\reference{} Schneider, D.P., van Gorkom, J.H,
Schmidt, M., \& Gunn, J.E. 1992, AJ 103, 1451 
\reference{} Schwartz, D.A., 2003, ApJ, in press., ASTROPH 0202190
\reference{} Sharp, R.G., McMahon, R.G., Irwin, M.J., \& Hodgkin,
S.T., 2001, MNRAS 326, L45 
\reference{} Shields, G.A., Gebhardt, K., Salviander, S. et
al. 2002, ApJ, in press 
\reference{} Shioya, Y., Taniguchi, Y.,
Murayama, T., Ajiki, M., et al. 2003, PASJ, in press, ASTROPH 021388
\reference{} Sopp, H. \& Alexander, P., 1991, MNRAS 251P 14
\reference{} Stern, D., Djorgovski, S.G., Perley, R.A., de Carvalho,
R.R, \& Wall, J.W. 2000b, AJ, 119, 1526 
\reference{}Stern, D.,
Spinard, H., Eisenhard, P., Bunker, A., et al. 2000a, ApJ, 533, L75
\reference{} Stocke, J., Morris, S. L., Weymann, R. J., \& Foltz, C. B. 1992, ApJ, 396, 487
\reference{} Tremaine, S., Gebhardt, K., Bender, R., Bower, G., et al. 2002, ApJ 574, 740 
\reference{} Tzioumis, T., de Vries, W., Snellen, I., Koekemoer,
A. 2002, {\sl The third GPS/CSS workshop,'},
Pub. Astron. Soc. Australia, Vol 20, (ASA: Sydney) 
\reference{} Visnovsky, K. L., Impey, C. D., Foltz, C. B., Hewett, P. C., et al. 1992, ApJ, 391, 560\reference{} York, D.G., Adelman, J., Anderson, J.E.,
Anderson, S.F., et al. 2000, AJ 120, 1579 
\reference{}Yun, M.S. \& Carilli, C. L. 2002, ApJ 568, 88 
\reference{}Yun, M.S., Reddy, N.A., \& Condon, J.J. 2001, ApJ 554, 803 
\reference{}Zheng, W., Tsvetanov,
Z., Schneider, D., Fan, X., et al. 2000, AJ 120, 1607 
\reference{}Zylka, R. 1998, {\sl MOPSI Users Manual}, (IRAM: Grenoble)
\end{references}
\end{document}